\documentclass[11pt]{article}

\usepackage[margin=1in]{geometry}

\usepackage{siunitx}
\usepackage{upgreek}
\usepackage{graphicx}
\usepackage{amsmath}
\usepackage{amssymb}
\usepackage{xfrac}
\usepackage{xcolor}
\usepackage{soul}

\usepackage{bm}

\usepackage{mathptmx}

\usepackage[super,comma,sort&compress]{natbib}

\usepackage{titlesec}
\titleformat{\section}{\normalsize\bfseries}{\thesection.}{1em}{}[]
\titleformat{\subsection}{\normalsize\bfseries}{\thesubsection.}{1em}{}[]
\titleformat{\subsubsection}{\normalsize\itshape}{\thesubsubsection.}{1em}{}[]
\titlespacing{\section}{0pt}{5pt}{0pt}[0pt]
\titlespacing{\subsection}{0pt}{3pt}{0pt}[0pt] 
\titlespacing{\subsubsection}{0pt}{0pt}{0pt}[0pt] 

\usepackage[font=footnotesize,labelfont=bf]{caption} 

\usepackage{float}
\usepackage[color=blue!15, size=footnotesize]{todonotes}
\usepackage{xcolor}

\begin{document}

\begin{center}

\Large
\textbf{Energy injection in an epithelial cell monolayer indicated by negative viscosity}

\vspace{11pt}

\normalsize
Molly McCord$^{1,2}$, Jacob Notbohm$^{1,2,*}$
\vspace{11pt}

$^1$Department of Mechanical Engineering, University of Wisconsin--Madison, Madison, WI, USA\\
$^2$Biophysics Program, University of Wisconsin--Madison, Madison, WI, USA

$^*$Contact author: jknotbohm@wisc.edu

\end{center}

\vspace{11pt}

\section*{ABSTRACT}

Epithelial tissues are driven out of thermodynamic equilibrium by internally generated forces, causing complex patterns of motion. Even when both the forces and motion are measurable, it is not yet possible to relate the two, because the sources of energy injection and dissipation are often unclear. Here, we study how energy is transferred by developing a method to measure the effective viscosity from the shear stresses and strain rates within an epithelial cell monolayer. Interestingly, there emerged multicellular regions in which the relationship between shear stress and shear strain rate was negatively proportional, indicating a negative effective viscosity. The negative effective viscosity occurred in regions wherein cell stresses were less efficient at producing tissue deformations compared to regions of positive effective viscosity. Regions of negative effective viscosity consistently exhibited greater cell speed and vorticity, and the cells had elevated metabolic activity, reflecting an increased energy demand in these cells. Our study shows that negative effective viscosity is a useful means of quantifying the flow of energy in living matter.

\section*{I. INTRODUCTION}

A central question in active matter is how the energy injected into the system is converted to forces, and, in turn, how those forces produce motion. In an epithelial cell layer, for example, the collective flows are  unpredictable, with rearrangements between neighboring cells and rotating eddies reminiscent of turbulence.\cite{poujade2007, vedula2012, alert2022review} Although the forces producing the flow can be measured experimentally, relationships between forces and flows remain complicated. Both cell-substrate tractions and cell-cell stresses fluctuate over space.\cite{trepat2009, tambe2011} In general, the direction of traction and the orientation of first principal stress do not align with the direction of cell motion.\cite{trepat2009, kim2013} As a result, some components of the tractions and stress inject energy to propel the flow, while other components act as a dissipative friction.\cite{bera2025} As the physical sources for energy injection and dissipation remain unclear, it is currently impossible to predict the cell motion from the forces.

Given that inertial forces are negligible in this system, it is commonly assumed that force and motion are related through a viscosity that couples shearing stresses to flows and dissipates the active forces produced by the cells.\cite{marchetti2013review, alert2020review}
No experiment has yet measured viscosity directly in this system. Order of magnitude estimates have suggested viscosity in cell collectives to be in a wide range, from $\sim$100 Pa$\cdot$hr to $\sim$10,000 Pa$\cdot$hr. \cite{duclos2017, perez2019} Numerical values for viscosity in this same range have also been inferred by fitting models to experimental data.\cite{marmottant2009, guevorkian2010, blanch2017} Prior efforts to measure viscosity directly by plotting shearing stresses against strain rates have failed, showing no clear trends,\cite{notbohm2016} leaving it unclear as to whether viscosity is a physically meaningful way to describe how the active forces produced by the cells are dissipated during the collective motion.

Here, we experimentally measured the effective viscosity within an epithelial cell layer by quantifying the ratio of shear stress to strain rate. In contrast to prior work,\cite{notbohm2016} we quantified the effective viscosity at different positions in space, thereby accounting for spatial variation in the magnitude and orientation of cell stresses. The results revealed a broad distribution of effective viscosity, with many values even becoming negative. The discovery of negative effective viscosity implies injection---rather than dissipation---of energy into the flow. We perturbed cell force production, finding that even when the flow rates were substantially altered, a substantial fraction of cells continued to exhibit negative effective viscosity. Finally, we revealed that cells with negative effective viscosity had elevated metabolism, meaning that our measurement of negative effective viscosity is a means of quantifying injection of energy into the flow.

\section*{II. RESULTS}

\subsection*{A. Negative Viscosity in an Epithelial Cell Monolayer}

\begin{figure}[t!]
\centering
\includegraphics[keepaspectratio=true, width=5.8in]{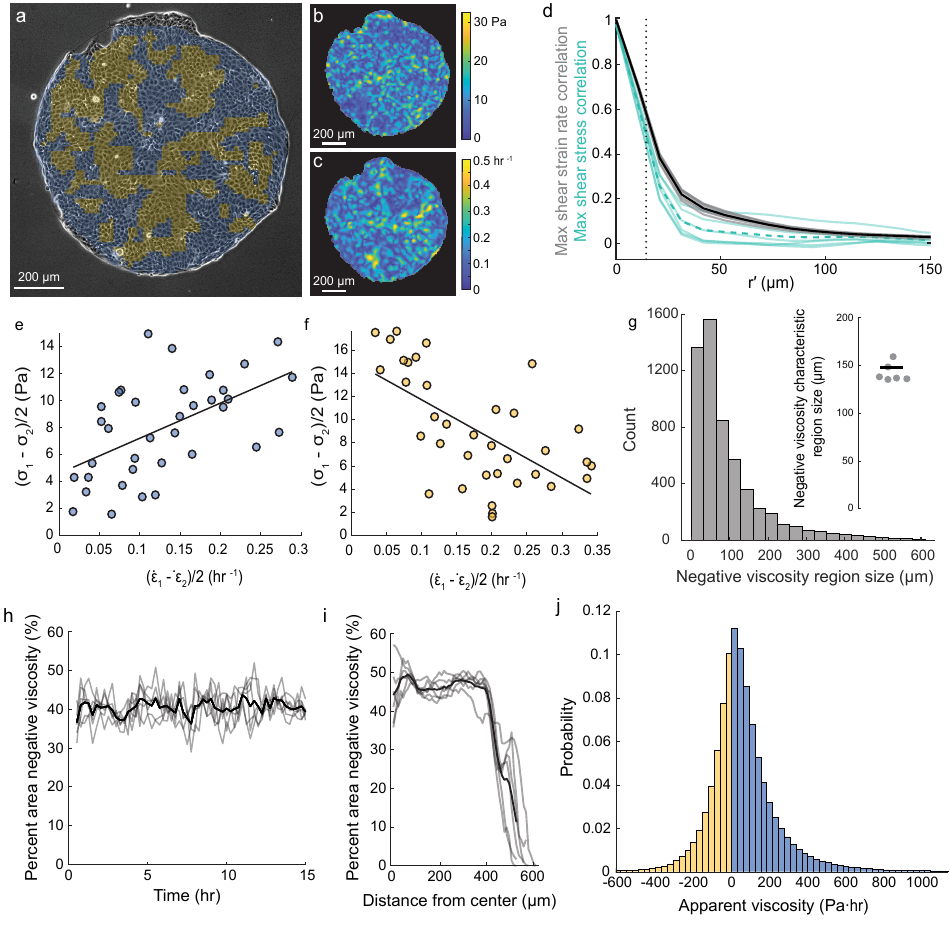}
\caption{Viscosity in the cell monolayer.
(a) Phase contrast image of a cell monolayer overlaid with colors indicating regions of positive (blue) and negative (yellow) viscosity. (b) Heat map of shear stress, $(\sigma_1 - \sigma_1)/2$. (c) Heat map of shear strain rate $(\dot{\varepsilon}_1 - \dot{\varepsilon}_2)/2$.
(d) Spatial correlation of shear strain rate (gray) and shear stress (turquoise). The solid and dashed lines indicate averages of the respective spatial correlations; the dotted line indicates the size of a single cell. 
(e) Scatter plot of shear stress and shear strain rate within a representative window showing positive viscosity (Pearson's correlation coefficient: 0.55). The slope of the line is $25.9 \pm 6.9$ Pa$\cdot$hr (slope $\pm$ standard error from linear regression).
(f) Scatter plot of shear stress and shear strain rate within a representative window showing negative viscosity (Pearson's correlation coefficient: -0.62). The slope of the line is  $-33.6 \pm 7.3$ Pa$\cdot$hr (slope $\pm$ standard error from linear regression). 
(g) Distribution of negative viscosity region size over 6 cell islands over 15 hr of imaging. Inset: Average size of regions of negative viscosity. 
(h) Percentage of the total island area exhibiting negative viscosity over time. 
(i) Percentage of the cells exhibiting negative viscosity against distance from the center of the island. 
(j) Histogram showing apparent viscosity over 6 cell islands over 15 hr of imaging. In panels g--i, gray dots and gray lines are data from different cell monolayers; black lines indicate means.}
\label{fig1}
\end{figure}

To quantify the effective viscosity, we began by measuring shear stress and shear strain rate in a Madin-Darby canine kidney (MDCK) cell monolayer, reasoning that their ratio would be the effective viscosity. To this end, we confined MDCK cells to 1 mm circular islands at a relatively low density of $\approx$2500 cells/mm$^2$ (Fig. 1a), which is low enough that the cells exhibit primarily fluid-like behavior.\cite{saraswathibhatla2020prx} Stresses within the cell monolayer were computed using Monolayer Stress Microscopy,\cite{tambe2011, tambe2013, saraswathibhatla2020sciData} and cell velocities were quantified using image correlation. The gradients of the velocities were computed, from which the components of the strain rate tensor $\dot{\bm{\varepsilon}}$ were calculated as described in the methods section. Next, we computed the eigenvalues of the stress tensor and strain rate tensor, which are referred to as the principal stresses, $\sigma_1$ and $\sigma_2$, and principal strain rates, $\dot{\varepsilon}_1$ and $\dot{\varepsilon}_2$. Here, we use the convention $\sigma_1>\sigma_2$ and $\dot{\varepsilon}_1 > \dot{\varepsilon}_2$. The average of the principal stresses is a tension (negative pressure), which tends to produce local changes in size; the shearing stresses are quantified by how different are the principal stresses, which is expressed mathematically as $(\sigma_1 - \sigma_2)/2$. Hence, we computed the shearing stress and strain rate according to $(\sigma_1 - \sigma_2)/2$ and $(\dot{\varepsilon}_1 - \dot{\varepsilon}_2)/2$, respectively. 
Representative color maps of shear stress and strain rate showed no obvious relationship between the two (Fig. 1b,c). Similarly, a scatter plot of all data points in the cell monolayer showed no correlation between the two (Appendix D, Fig. \ref{shear_island}), consistent with prior observations.\cite{notbohm2016} 

To investigate more deeply, we considered that there may be variability over space caused by the spatial variations in cell stresses. To quantify spatial variability, we performed spatial autocorrelations of shear stress and strain rate, which decayed on average over distances of 17.5 {\textmu}m and 22.7 {\textmu}m, respectively (Fig. 1d). These correlation lengths are about equal to the size of a cell. To account for this variability in space, we collected the data within a 62$\times$62 {\textmu}m$^2$ window (see the Methods section for an explanation of the choice of window size) and plotted shear stress against shear strain rate. We then performed a linear least-squares regression, and used the slope of the best fit line as the effective viscosity for that window. The data showed a clear linear correlation, with the slope indicating the effective viscosity (Fig. 1e). Interestingly, in a different randomly chosen window, the linear correlation was negative, indicating the existence of a negative effective viscosity (Fig. 1f). The observation of negative effective viscosity indicates that the active shear stresses produced by the cells are larger than the passive viscous stresses, meaning that the active shear stresses inject energy into the flowing cell layer.
This finding is reminiscent of theory and experiments in other systems, for which activity can reduce the effective viscosity.\cite{hatwalne2004, cates2008, sokolov2009, gachelin2013, lopez2015, orihara2019, chui2021} We note that this measurement is an \textit{effective} viscosity, resulting from both dissipation of energy, as occurs by viscosity in a passive fluid, and the active stresses produced by the cells. For brevity, hereafter we drop the word ``effective'' and simply use the term viscosity.

To build on our initial observations, we moved the window to all locations in the cell island, calculating the ratio of shear stress to strain rate at all locations in space. To increase the spatial resolution of the measurement, we overlapped all windows, such that the spacing between window centers matched the spatial resolution of the measurements of stress and strain rate, 10.4 {\textmu}m. This approach revealed distinct regions with positive (blue) and negative (yellow) viscosity (Fig. 1a). To quantify the characteristic size of a region with negative viscosity, we computed the square root of the area of each connected region having negative viscosity and plotted the distribution as a histogram (Fig. 1g). The mean of the distribution was also plotted for six different cell islands, and the average was 149 {\textmu}m (9--10 cell widths) (Fig. 1g).
We also verified that the choice of window size did not substantially affect the typical size of regions of negative viscosity. To this end, we repeated the analysis, quantifying the characteristic size of a region with negative viscosity for a smaller and larger window. The results showed that the characteristic size was relatively insensitive to window size (Appendix D, Fig. \ref{window}).

In addition to a shear viscosity, which quantifies resistance to local changes in shape, there can also be a bulk viscosity, which quantifies resistance to local changes in area. To check for the existence of a bulk viscosity, we used the average principal stress and strain rate, $(\sigma_1 + \sigma_2)/2$ and $(\dot{\varepsilon}_1 + \dot{\varepsilon}_2)/2$, which showed similar patterns of negative and positive values (Appendix D, Fig. \ref{bulk_viscosity}).
Given that motion within these confined islands requires cells to undergo shearing (shape changing) deformations to slide past one another, we focused on the shear viscosity for the remainder of our study. Next, to demonstrate that negative shear viscosity is a general phenomenon that exists in monolayers of other cell types, we repeated these experiments with the human keratinocyte cell line HaCaT, which exhibited negative viscosity as well (Appendix D, Fig. \ref{hacat}).

To explore the temporal evolution of regions of negative viscosity, we measured the percentage of the total area within the cell island exhibiting negative viscosity over time, with results showing that for all times, $\approx$40\% of the cell island exhibited negative viscosity (Fig. 1h). We also determined how negative viscosity depended on position, with the locations near the edge of the island tending to exhibit positive viscosity (Fig. 1i). To avoid possible artifacts associated with boundaries, we analyzed data in locations $>100$ {\textmu}m from the boundaries for the rest of our study.
Finally, we plotted a histogram of the signed value of viscosity for all locations in 6 different cell islands, with results ranging from -600 to 800 Pa$\cdot$hr and the distribution centered on values slightly greater than 0 (Fig. 1j). In a separate study, we have shown that the positive values of viscosity are related to the cell cytoskeleton and cell-cell adhesions,\cite{mccord_positiveviscosity} and here we focus on negative values of viscosity.

Although a reduction in viscosity due to activity has been observed in other systems,\cite{sokolov2009, gachelin2013, lopez2015, orihara2019, chui2021} a distinction here is that the prior studies measured the average viscosity of the system, whereas here the data quantify the apparent viscosity at every point in space. Importantly, the finding that negative viscosity occurs in regions spanning many cell widths indicates that the negative viscosity is unlikely to result from random experimental errors. Moreover, the regions of negative viscosity are not caused by individual cells or individual rearrangement events; rather, the data suggest the existence of a more complex intercellular coordination.

\subsection*{B. Orientations of Stresses and Strain Rates in Regions of Negative and Positive Viscosity}

\begin{figure}[t!]
\centering
\includegraphics[keepaspectratio=true, width=6.5in]{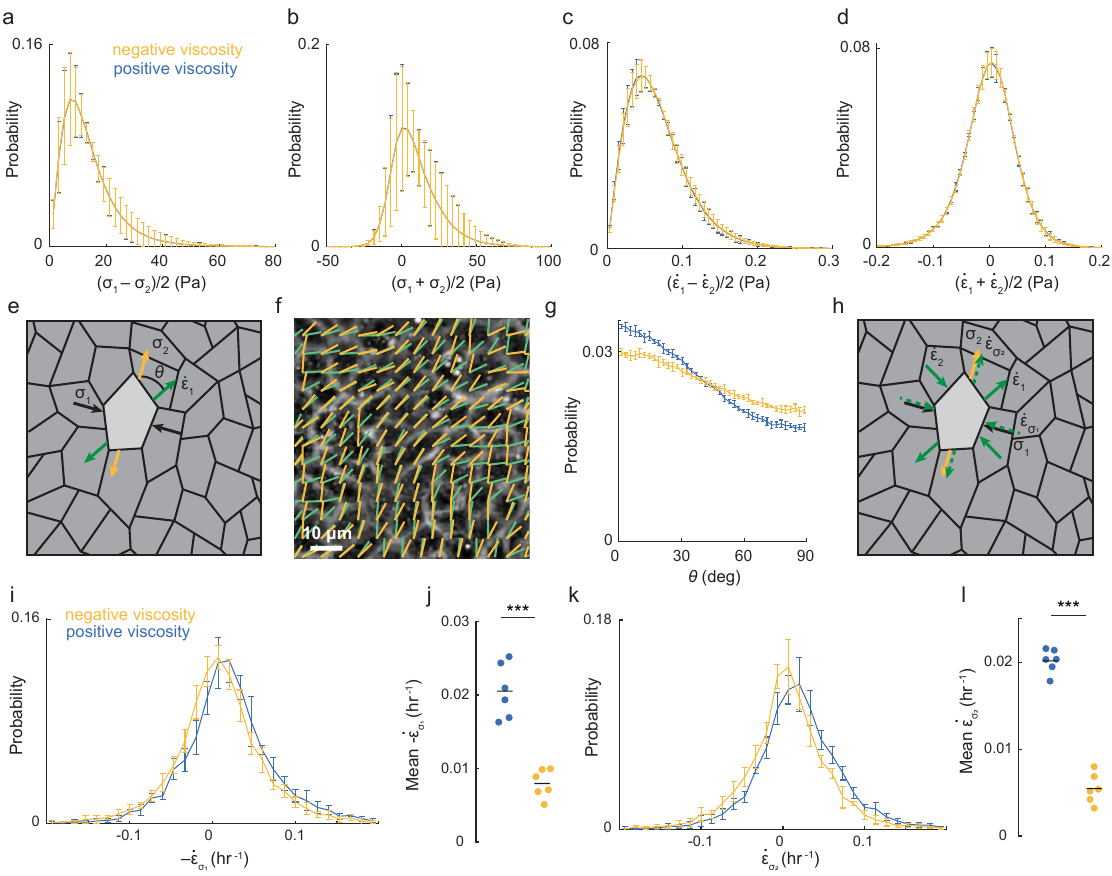}
\caption{Stresses and strain rates in regions of positive and negative viscosity. 
(a) Histogram of $(\sigma_1-\sigma_2)/2$ in regions of positive and negative viscosity. 
(b) Histogram of $(\sigma_1+\sigma_2)/2$ in regions of positive and negative viscosity. 
(c) Histogram of $(\dot{\varepsilon}_1 - \dot{\varepsilon}_2)/2$ in regions of positive and negative viscosity. 
(d) Histogram of $(\dot{\varepsilon}_1 + \dot{\varepsilon}_2)/2$ in regions of positive and negative viscosity. 
(e) Depiction of the stresses (black and yellow arrows) applied by a cell (light gray) onto its neighbors (dark gray). Negative viscosity would occur if the orientation of the second principal stress, $\sigma_2$ (yellow arrows), aligns with the orientation of the first principal strain rate, $\dot{\varepsilon}_1$ (green arrows). The difference between orientations of $\sigma_2$ and $\dot{\varepsilon_1}$ is indicated by angle $\theta$.
(f) Image of a negative viscosity region overlaid with lines depicting the orientations of $\sigma_2$ (yellow) and $\dot{\varepsilon}_1$ (green).
(g) Histogram of angle $\theta$ between orientations of $\sigma_2$ and $\dot{\varepsilon}_1$ in regions of negative and positive viscosity ($p < 0.0001$, Kolmogorov-Smirnov test comparing the two distributions).
(h) Depiction of the stresses (black and yellow arrows) applied by a cell (light gray) onto its neighbors (dark gray). The principal strain rates are shown by the solid green arrows. The strain rate tensor was projected into orientation of stress, and the dashed arrows show the diagonal components of the projected strain rates, $\dot{\varepsilon}_{\sigma_1}$ and $\dot{\varepsilon}_{\sigma_2}$.
(i) Distribution of $-\dot{\varepsilon}_{\sigma_1}$ in well-defined regions of positive and negative viscosity.
(j) Mean of $-\dot{\varepsilon}_{\sigma_1}$ for well-defined regions of positive and negative viscosity ($p < 0.0001$, two-sample t-test). Each dot represents a cell island.
(k) Distribution of $\dot{\varepsilon}_{\sigma_2}$ in well-defined regions of positive and negative viscosity.
(l) Mean of $\dot{\varepsilon}_{\sigma_2}$ for well-defined regions of positive and negative viscosity ($p < 0.0001$, two-sample t-test). Each dot represents a cell island.
All histograms are over 6 cell islands and 15 hr of imaging; error bars represent the standard deviation of the probability values for each bin, calculated across all cell islands.}
\label{fig2}
\end{figure}

To develop an explanation for how the viscosity could become negative, we next studied the stresses and strain rates more closely. We began by quantifying distributions of stresses in regions of positive and negative viscosity. The shear stress, $(\sigma_1-\sigma_2)/2$, ranged from 0 to 80 Pa with no notable difference between distributions in regions of positive or negative viscosity (Fig. 2a). The average normal stress, $(\sigma_1+\sigma_2)/2$ had a nearly symmetric distribution shifted toward positive values, indicating a tendency for tensile (pulling) stresses, again with no difference between regions of positive and negative viscosity (Fig. 2b). We also analyzed the shear strain rate, $(\dot{\varepsilon}_1-\dot{\varepsilon}_2)/2$, and average normal strain rate, $(\dot{\varepsilon}_1+\dot{\varepsilon}_2)/2$ within these regions. Similar to stresses, the distributions of strain rate showed no notable difference between regions of positive and negative viscosity (Fig. 2c,d).

Next, we considered the explanation from theories developed for active systems such as swimming bacteria and active nematic materials,\cite{hatwalne2004, cates2008, saintillan2018} which showed that if activity creates pushing (extensile) stresses that locally align with the predominant orientation of extensile flow, the effective viscosity can be reduced and even become negative. Following our sign convention, the largest pushing stress is the second principal stress $\sigma_2$. For $\sigma_2$ to propel the flow, it would have to align with the first principal strain rate $\dot{\varepsilon}_1$ (Fig. 2e). Thus, we plotted the orientations of $\sigma_2$ and $\dot{\varepsilon}_1$ in a representative field of view, with results qualitatively showing a tendency for alignment (Fig. 2f).

To quantify the alignment, we plotted histograms of angle $\theta$ between the orientations of $\sigma_2$ and $\dot{\varepsilon}_1$. Although slight, the histograms exhibited a peak at zero, indicating a tendency for alignment between $\sigma_2$ and $\dot{\varepsilon}_1$ (Fig. 2g). Importantly, the peak at zero was statistically meaningful, as the variability in the histograms between different islands was negligible (Fig. 2g). Unexpectedly, the tendency for alignment was greater for regions of positive viscosity compared to negative. To ensure that this trend was not driven by transitional zones at the boundaries between the regions of positive and negative viscosity, we repeated the analysis using only well-defined regions of positive and negative viscosity, which were chosen by excluding data within 31 {\textmu}m (equal to half the width of the windows used to quantify viscosity) of the edge of each region; the resulting histograms of $\theta$ (Appendix D, Fig. \ref{extensile_contractile}) were essentially the same as in Fig. 2g.

To investigate further, we considered the recent finding that both extensile and contractile behavior occur within the same cell island,\cite{nejad2024} where ``contractile'' refers to cells whose body tends to align with the orientation of maximal principal stress, $\sigma_1$, and, ``extensile,'' cells that align with the orientation of $\sigma_2$. Histograms of angle $\theta$ for only contractile and only extensile cells showed the same trends as the histograms for all cells, namely, there was a slight tendency of $\theta$ towards 0, and the tendency was stronger in regions of positive viscosity compared regions of negative viscosity (Appendix D, Fig. \ref{extensile_contractile}). We next asked whether the tendency for alignment between cell stresses and the flow depended on cell aspect ratio. We identified elongated cells (see Methods) and plotted histograms of angle $\theta$, which showed an approximately equal tendency toward zero for cells in regions of positive and negative viscosity (Appendix D, Fig. \ref{extensile_contractile}). 

In summary, the results were only partially consistent with theoretical predictions. The observation that $\theta$ tends toward zero supports the idea of alignment between orientations of $\sigma_2$ and $\dot{\varepsilon}_1$, as expected. However, the greater alignment of $\theta$ in regions of positive viscosity than in regions of negative viscosity was unexpected. For this reason, we considered an alternative approach of measuring the effective viscosity. The alternative approach did not use a window but instead transformed the stress tensor into the orientation of principal strain rates and directly compared stresses and strain rates in that orientation (Appendix A). However, our analysis showed that, if defined based on this approach, regions of negative viscosity were not meaningfully related to energy injection within the monolayer (Appendix A and Fig. 10). This result prompted us to return to the method established in Fig. 1 and to reconsider whether orientation alone could adequately distinguish between regions of positive and negative viscosity.

We next asked whether the deformation magnitudes, rather than the orientation alone could better distinguish between regions of positive and negative viscosity. In our system, the orientation of maximal principal stress, $\sigma_1$, corresponds to the orientation that the largest contractile forces are generated. We focused on how much deformation occurred along this contractile axis. To assess this, we projected the strain rate tensor onto the orientation of $\sigma_1$ and computed $-\dot{\varepsilon}_{\sigma_1}$, the component of strain rate aligned with the orientation of $\sigma_1$, as illustrated in Fig. 2h. (The negative sign is used here, because if the cell were isolated, it would get smaller along this axis, meaning the strain would be negative.) We then plotted a histogram of $-\dot{\varepsilon}_{\sigma_1}$ over all time points and all cell islands in regions of positive and negative viscosity (Fig. 2i). For this analysis, we focused on well-defined regions of positive or negative viscosity by excluding data within 31 {\textmu}m  of the edge of each region, and each data point in the histogram is an average over a well-defined region. In contrast to maximum shear strain rate, whose mean was nearly identical in regions of positive and negative viscosity (Fig. 2c), the  mean of $-\dot{\varepsilon}_{\sigma_1}$ over all cell islands was approximately threefold lower in regions of negative viscosity regions compared to positive viscosity (Fig. 2j). These results suggest that regions of negative viscosity had limited deformation in the direction of contraction; in contrast, positive viscosity regions exhibited more substantial tissue deformation.

Given that stress is distributed along both principal axes, we next examined whether similar deformation patterns occurred along the orientation of $\sigma_2$. We transformed the strain rate tensor into the orientation of $\sigma_2$ to compute $\dot{\varepsilon}_{\sigma_2}$ for regions of positive and negative viscosity. We then plotted the distribution of $\dot{\varepsilon}_{\sigma_2}$ over all islands and time in regions of positive and negative viscosity (Fig. 2k). Similar to previous results, the mean of strain rate across islands was approximately 3.5-fold lower in negative viscosity regions compared to positive viscosity regions (Fig. 2l). Together, these results suggest that in regions of negative viscosity, tissue deformations were smaller along the orientations of the principal stresses. Hence, the explanation of negative viscosity in other systems, based on alignment of stress and strain rate,\cite{sokolov2009, gachelin2013, lopez2015, orihara2019} only partly describes the behavior observed in epithelial cell monolayers.

\subsection*{C. Negative Viscosity is Robust to Perturbations to Stresses and Flow Rates}

Next, we asked whether we could perturb the fraction of cells exhibiting negative viscosity by increasing or decreasing activity in the cell monolayer. We used two treatments, CN03 and cytochalasin D, which increase and decrease force generation, respectively.\cite{saraswathibhatla2020prx} In these experiments, we imaged the cell islands in control conditions for 1 hr, then treated with vehicle control, CN03, or cytochalasin D and imaged for an additional 14 hr. Representative images of shear stress and strain rate at 1.5 hr after treatment are shown in Fig. 3a. Results from multiple different experiments showed that CN03 increased the shear stress and strain rate by factors of approximately 2 and 1.5, respectively, whereas cytochalasin D decreased the shear stress and strain rate by approximately 20\% and 50\%, respectively (Fig. 3b, c). 

Next, we determined locations in the cell islands having positive and negative viscosity at each time point under the CN03 and cytochalasin D treatment (Fig. 3a). We then quantified the fraction of area having negative viscosity. Despite the notable changes in stress and strain rate caused by the treatments, the fraction of the island area having negative viscosity was unchanged by the treatments, being approximately 40\% of the cell monolayer in all conditions (Fig. 3d).

\begin{figure}[t!]
\vspace{-20pt plus 2pt minus 2pt}
\centering
\includegraphics[keepaspectratio=true, width=6.1in]{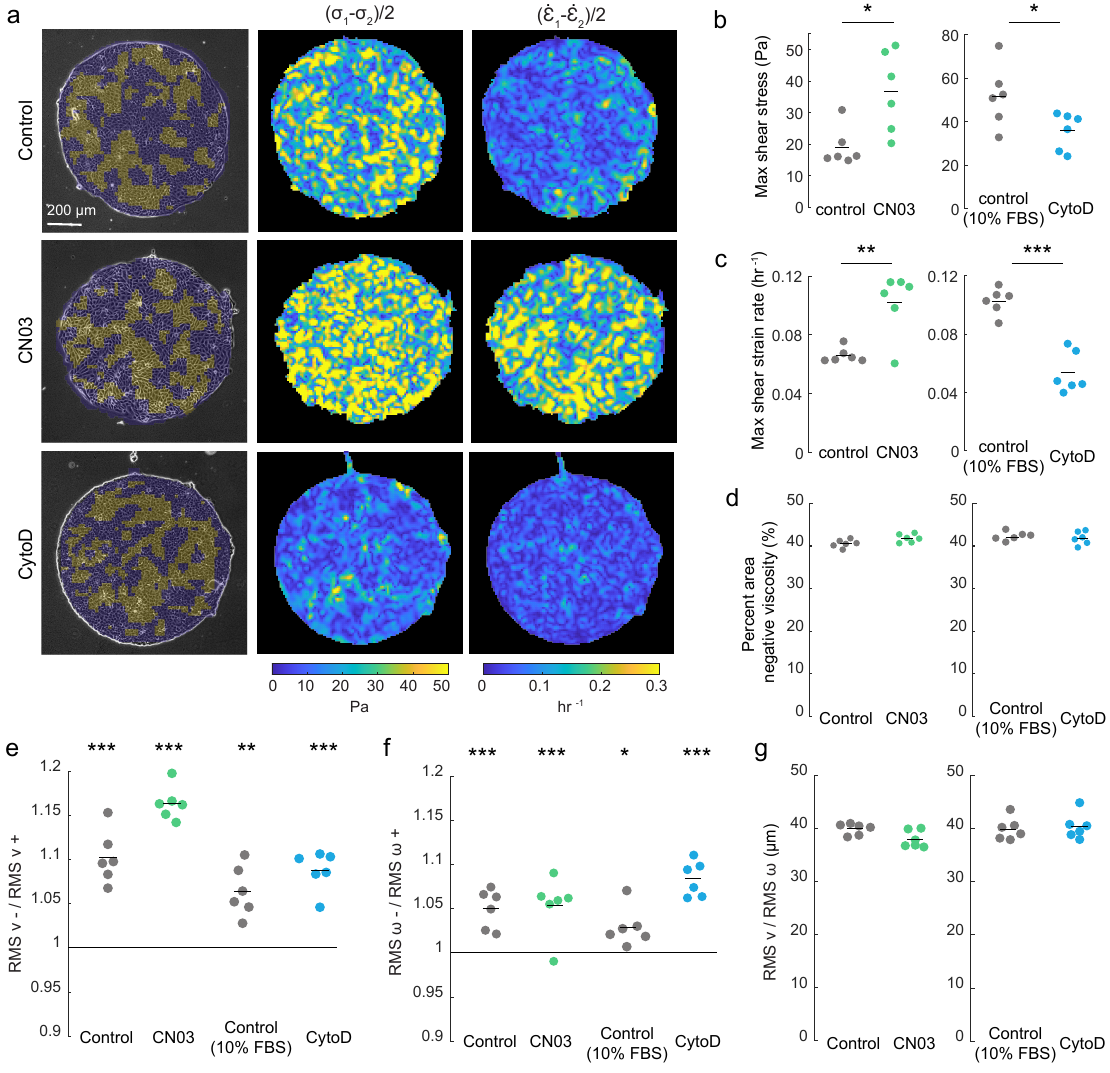}
\vspace{-15pt plus 2pt minus 2pt}
\caption{Altering stresses and flow rates does not alter the fraction of cells with negative viscosity. 
(a) Phase contrast image of a cell monolayer (left) overlaid with regions of positive (blue) and negative (yellow) viscosity. Heat maps of maximum shear stress (center) and maximum shear strain rate (right). Data are shown for control cells (top row) and cells treated with CN03 (middle row) and cytochalasin D (cytoD, bottom row). 
(b) Shear stress for treatments and their respective vehicle controls. (CN03 $p = 0.012$, cytoD $p = 0.039$, two-sample t-tests). As described in methods, the control for cytoD used a higher concentration of fetal bovine serum (FBS), 10\%.
(c) Shear strain rate for treatments and their respective vehicle controls (CN03 $p =  0.002$, cytoD $p < 0.0001$, two-sample t-tests).
(d) Percent negative viscosity within the monolayer for control and treatment cases (CN03 $p = 0.5$, cytoD $p = 0.99$, two sample t-tests).
(e) Ratio of RMS velocity in regions of negative viscosity to positive viscosity for control and treatment cases (control $p < 0.0001$, CN03 $p < 0.0001$, control 10$\%$ FBS $p = 0.002$, cytoD $p < 0.0001$, one-sample t-tests in comparison to 1). 
(f) Ratio of RMS vorticity, $\omega$, in regions of negative viscosity to positive viscosity for control and treatment cases (control $p < 0.0001$, CN03 $p < 0.0001$, control 10$\%$ FBS $p = 0.02$, cytoD $p < 0.0001$, one-sample t-tests in comparison to 1).
(g) Characteristic size of a vortex, given by RMS v / RMS $\omega$, for the control and treatment cases (CN03 $p = 0.11$, cytoD $p = 0.09$, two sample t-tests). For all panels, a dot indicates the mean of a cell island over time, and black bars indicate means. }
\label{fig3}
\end{figure}

Following the observation that the fraction of the cells exhibiting negative viscosity was unaffected by increasing or decreasing stresses and strain rates, we considered how negative viscosity affected the cell flow. We predicted that, given negative viscosity indicates energy injection by active shearing stresses into the flow, cells in regions of negative viscosity would move faster compared to cells in regions of positive viscosity. To begin, we ruled out an alternative explanation that, in regions of negative viscosity, energy injected via cell-substrate traction might be reduced (Appendix B and Appendix D, Fig. \ref{traction_velocity}). Next, we computed the root-mean-square (RMS) of cell velocity in regions of negative and positive velocity and took the ratio. Results indicated that in all conditions, the average ratio was $>1$, typically in the range 1.05--1.2 (Fig. 3e). Hence, in regions of negative viscosity, cells move faster than in regions of positive viscosity. Current understanding is that viscous energy dissipation at the substrate is proportional to velocity,\cite{lee2011, cochet2014, notbohm2016, blanch2017, perez2019, alert2020review, vazquez2022} meaning for cells that inject energy, as indicated by negative viscosity, that surplus energy is likely dissipated at the cell-substrate interface.

We also studied the vorticity $\omega$ in regions of negative and positive viscosity, with results being similar, namely vorticity was elevated in regions of negative viscosity (Fig. 3f). 
Finally, wondering whether negative viscosity is related to the common observation of rotational eddies in the cell monolayer reminiscent of turbulence,\cite{poujade2007, vedula2012, alert2022review} we quantified the characteristic size of a turbulent eddy by computing the ratio of the  RMS of velocity divided by the RMS of vorticity. This analysis resulted in a characteristic size of approximately 40 {\textmu}m under all conditions (Fig. 3g), which is similar to a prior study\cite{lin2021} but substantially different than the typical size of a region having negative viscosity ($\approx$150 {\textmu}m, Fig. 1g), suggesting that our observation of negative viscosity is distinct from the common observation of turbulent-like rotation within the flow.

\subsection*{D. Source of Energy for Negative Viscosity}

The injection of energy as indicated by negative viscosity, combined with the observation of faster cell speeds in regions of negative viscosity, raise the question of whether cells in regions of negative viscosity produce energy at an elevated rate. We used fluorescent imaging of glucose uptake, via (2-(N-(7-Nitrobenz-2-oxa-1,3-diazol-4-yl)Amino)-2-Deoxyglucose (2-NBDG), and mitochondrial membrane potential, via Tetramethylrhodamine, Ethyl Ester, Perchlorate (TMRE), both of which are markers of metabolic activity\cite{scaduto_measurement_1999, zou2005, decamp2020} (Fig. 4a, b). To quantify the results, we determined the average fluorescent intensity in regions of negative and positive viscosity, and computed the ratio for multiple different cell islands. As above, we excluded regions within 100 {\textmu}m of the boundary of the cell monolayer. The data consistently showed that regions of negative viscosity displayed both elevated glucose uptake and elevated mitochondrial membrane potential (Fig. 4c), indicating increased metabolic activity in regions of negative viscosity. Additionally, the florescent intensity of both glucose uptake and mitochondrial membrane potential was smaller at the edges of the cell island compared to the center (Fig. 4d), matching the observation that the fraction of area having negative viscosity was also smaller on the edges of the cell island (Fig. 1i).

Following our prior measurement that a region of negative viscosity has characteristic size of 149 {\textmu}m (Fig. 1g), we also computed spatial autocorrelations of the images of both glucose uptake and mitochondrial membrane potential (Fig. 4e). The autocorrelations decayed sharply within the first 20 {\textmu}m, which is due to the fluorescent imaging clearly defining the cell outlines (Fig. 4a, b insets). After the first 20 {\textmu}m, the decay was slower, indicating that glucose uptake and mitochondrial membrane potential were spatially correlated, possibly due to intercellular coordination. The spatial correlation lengths of glucose uptake and mitochondrial membrane potential were 118 and 123 {\textmu}m, which is close to the characteristic size of a region of negative viscosity ($\approx$150 {\textmu}m, Fig. 1g). 

\begin{figure}[t!]
\vspace{-20pt plus 2pt minus 2pt}
\centering
\includegraphics[keepaspectratio=true, width=6in]{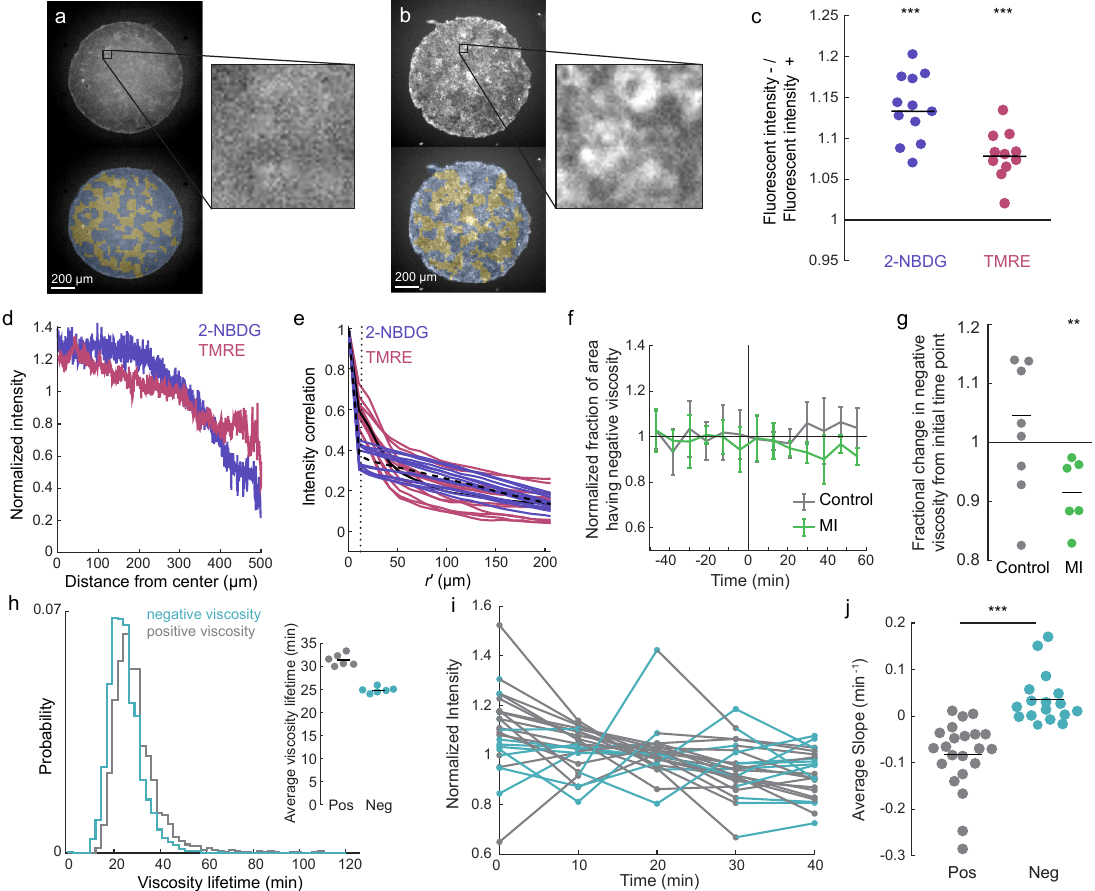}
\vspace{-3pt plus 2pt minus 2pt}
\caption{Regions of negative viscosity are more metabolically active. 
(a) Fluorescent image of cell island stained with 2-NBDG (top) and overlaid with map depicting positive (blue) and negative (yellow) viscosity (bottom). The black box is an enlarged view, showing the outlines of individual cells. 
(b) Fluorescent image of cell island stained with TMRE (top) and overlaid with map depicting positive (blue) and negative (yellow) viscosity (bottom). The black box is an enlarged view, showing the outlines of individual cells. 
(c) Ratio of 2-NBDG and TMRE fluorescent intensity in regions of negative viscosity to regions of positive viscosity (2-NBDG $p < 0.0001$, TMRE $p < 0.0001$, one-sample t-tests in comparison to 1). 
(d) Fluorescent intensity of 2-NBDG and TMRE was normalized such that the average was unity for each cell island and then plotted against distance from the center. The lines show data averaged over 12 islands treated with 2-NBDG and 11 islands treated with TMRE.
(e) Spatial correlation of TMRE and 2-NBDG fluorescent intensity. The dotted line indicates the size of a single cell. 
(f) Normalized fraction of area having negative viscosity before and after metabolic inhibition (MI). All values were normalized by the mean of times $\in (-50,0)$ min. The metabolic inhibition occurred at $t=0$ min. Lines represent the mean and error bars represent the standard deviation. 
(g) Fractional change in area of negative viscosity from initial time point to last time point (Control $p = 0.396$, MI $p = 0.004$, one-sample t-test in comparison to 1). 
(h) Distribution of viscosity lifetime in regions of positive and negative viscosity. Inset: Average lifetime of regions of positive and negative viscosity. Each dot represents a different cell island.
(i) Fluorescent intensity of TMRE over time. Each line indicates a randomly chosen $62 \times 62$ {\textmu}m$^2$ region of interest (ROI). Gray and blue line segments indicate times over which the ROI had positive and negative viscosity, respectively. Fluorescent intensity in each ROI was normalized by the mean over all time points in that ROI. The sign of viscosity was determined based on the sign of the majority of grid points within the ROI at each time point. 
(j) Average slope of normalized fluorescent intensity for positive and negative viscosity line segments of each ROI ($p < 0.0001$, paired sample t-test). 
Dots in panels c, f, and g indicate averages over a cell island. Black bars indicate means.}
\label{fig4}
\end{figure}

The correlation between regions of elevated metabolism and negative viscosity led us to hypothesize that there may be a causal relationship wherein surplus energy is injected into the flow via active shear stresses, in turn causing negative viscosity. The closely matched sizes of regions of elevated metabolism and negative viscosity suggest that cells having surplus energy inject it into the flow using active shearing stresses. Before testing this hypothesis, we first ruled out an alternative explanation based on substrate-to-cell tractions, showing that cells with elevated metabolism do not inject energy into the cell layer via tractions (Appendix C, Appendix D, Fig. \ref{metabolism_phi}). Next, as a first test of our hypothesis, we imaged the cells over time before and after inhibiting metabolism. Both oxidative phosphorylation and glycolysis were inhibited, using NaCN and 2-Deoxy-d-glucose (2-DG), respectively. The two metabolic pathways were inhibited simultaneously to prevent, as much as possible, compensatory effects wherein inhibiting one pathway leads to compensation by the other. Importantly, under this treatment, metabolic activity decreased, but the cells were alive over the entire time course of our experiment, as indicated by the cell stresses and strain rates, which never reached zero (Appendix D, Fig. \ref{force_motion_metabolism_inhibition}). To quantify the effects of inhibiting metabolism on negative viscosity, for each cell island, we computed the area having negative viscosity at all time points and then normalized by the average area for time points before inhibition. In control experiments, the normalized area of negative viscosity was unchanged, maintaining a value of unity. By contrast, for cell islands with metabolism inhibited, the normalized area of negative viscosity decreased approximately 30 min following the inhibition (Fig. 4f). To quantify these results further, we averaged the normalized area of negative viscosity for all time points $>30$ min after treatment (chosen because it takes this long for the treatment to take effect in this cell type\cite{baron2005}). Whereas the fraction of cells having negative viscosity was unchanged in control conditions, the metabolic inhibition reduced the fraction of cells having negative viscosity by approximately 10\% (Fig. 4g). This 10\% reduction is comparable to the magnitude of metabolic suppression observed in previous studies following metabolic inhibition.\cite{walsh2021,filon2022} Hence, inhibiting metabolism reduced the fraction of cells exhibiting negative viscosity. 

To test further the hypothesis that elevated metabolism causes negative viscosity, we considered how both negative viscosity and metabolism vary over time. We first quantified the lifetime of regions of positive and negative viscosity (see Methods), which averaged 32 and 25 min, respectively (Fig. 4h). We then performed an experiment to image metabolism over time periods longer than this characteristic time. The fluorescent indicators of metabolism photobleach quickly, especially 2-NBDG, which photobleached too quickly for time lapse imaging. Fortunately, the imaging of mitochondrial membrane potential by TMRE was relatively stable when imaging every 10 min for 40 min, which is slightly longer than the characteristic lifetimes of regions of positive and negative viscosity.  We analyzed $62 \times 62$ {\textmu}m$^2$ regions of interest (ROIs), which were chosen randomly from 6 different islands. In each ROI, we quantified the average fluorescent intensity and determined whether the viscosity was negative or positive. We normalized the fluorescent intensity by the average for all time points and plotted the normalized intensity over time, with line color indicating positive or negative viscosity. The common trend in the data was that when the mitochondrial membrane potential decreased over time, the viscosity was positive, and when membrane potential increased over time, the viscosity was negative (Fig. 4i and Appendix D, Fig. \ref{4i_split}). To quantify this relationship, for the data in ROIs shown in Fig. 4i, we calculated the slope of each line segment in each 10 min time window between imaging. In each ROI, the slopes of all line segments in time windows corresponding to positive viscosity in that ROI were averaged, as were the slopes of all line segments corresponding to negative viscosity. The averaged slopes were then plotted for positive and negative viscosity, with results showing that slopes during periods of negative viscosity were significantly larger than those during periods of positive viscosity (Fig. 4j). Hence, these data indicate a temporal correlation, wherein increases in mitochondrial membrane potential tend to occur concurrently with negative viscosity. Together, these data suggest that cells that have surplus energy available inject that excess energy into the flow by active shearing stresses, as quantified by our measurement of negative viscosity.

\section*{III. DISCUSSION}

Our study was motivated by the fact that it is often unclear how energy is injected and dissipated in active biological materials. Given that viscosity indicates energy dissipation in passive fluids, we developed a method to quantify the effective viscosity over space and time within an epithelial cell monolayer. Interestingly, the results revealed regions of negative viscosity, resulting from the active stresses produced by the cells. Prior experiments in other systems demonstrated that active stresses can reduce the effective viscosity,\cite{sokolov2009, gachelin2013, lopez2015, orihara2019, chui2021} and the proposed physical mechanism was alignment between the orientations of second principal stress and first principal strain rate.\cite{hatwalne2004, cates2008, saintillan2018} Our experiments directly measured orientations of stresses and strain rates, and the results were partly consistent with the theoretical explanation in that there was a tendency for alignment between orientations of second principal stress and first principal strain rate. An unexpected result was that this alignment was stronger in regions of positive viscosity compared to negative, indicating orientation alone did not fully describe the mechanism for negative viscosity. Further analysis showed that the average magnitude of strains, projected into the orientations of the principal stresses, were substantially smaller in regions of negative viscosity compared to positive. Conceptually, this observation indicates that in regions of negative viscosity, the stresses were less efficient in producing tissue deformations. Finally, we showed that cells in regions of negative viscosity had elevated metabolism, indicating that our measurement of negative viscosity is a useful indicator of energy injection in epithelial cell monolayers.

The fact that prior theoretical explanations\cite{hatwalne2004, cates2008, saintillan2018} do not fully match our observations does not invalidate the theories, because the theories were developed for different experimental systems, namely swimming bacteria and active nematics. Monolayers of epithelial cells, as we study here, are highly complicated. For example, the first principal stress is often misaligned with the cell body,\cite{nejad2024} and, in addition to stresses, cell-substrate tractions strongly affect the collective motion.\cite{saraswathibhatla2020prx, saraswathibhatla2021, bera2025} It will be interesting to adapt the theories by incorporating these observations, which will enable a more precise understanding of the causes and implications of negative viscosity. Another interesting direction for future theoretical developments could come from the field of solid mechanics, wherein composite materials having inclusions of negative effective stiffness can be designed such that their modulus is arbitrarily large.\cite{lakes2002, jaglinski2007}

An advantage of our experiments that quantified viscosity in both space and time is that the data showed that regions of negative viscosity are spatially correlated, with a characteristic size of approximately 150 {\textmu}m. The regions of negative viscosity coincided with regions of elevated cellular metabolism. In contrast, metabolism was uncorrelated with regions wherein the cell-substrate tractions propelled the flow. Together, these observations suggest that cells having surplus energy may inject that energy into the flow by active shearing stresses, which causes the effective viscosity to become negative. Some prior models have used a negative viscosity term,\cite{wensink2012, rossen2014} which was intended as a simplified means of injecting energy into the system. Our observation of negative viscosity shows that a model using negative viscosity is, in fact, an accurate, albeit coarse grained, description of an epithelial cell monolayer. One point to emphasize, however, is that our data, especially the data showing negative viscosity coincides with elevated metabolism, show that any modeling approach should inject energy in spatial regions of characteristic size matching those measured here.

Interestingly, the fraction of the cell layer exhibiting negative viscosity was $\approx$40\% for nearly all experimental conditions, even in response to treating with CN03 and cytochalasin D, which substantially increased and decreased stresses and strain rates. This observation hints that there may be some underlying cellular process for achieving such robust behavior. It is reasonable to suspect that the robustness may be caused by cellular metabolism, which results from a highly complex network of interactions that create compensatory mechanisms which enable the cell to maintain a consistent metabolic activity.\cite{smart2008, stelling2004, whitacre2012} Consistent with this reasoning, only the treatment that directly inhibited metabolism had an effect on the fraction of cells exhibiting negative viscosity. Given that negative viscosity indicates energy injection by active shear stresses, it appears that it may have been evolutionarily favorable for cells to leverage shear stresses as a means to inject surplus energy into the collectively flowing monolayer. Implications of this finding could be far reaching. In metastasis, for example, cancer cells often have altered metabolic activity,\cite{deberardinis2016, bergers2021} which raises the question of whether the fraction of cells exhibiting negative viscosity is altered during cancer invasion. Other examples wherein metabolism is elevated include wound healing \cite{wang2024skinwound} and tissue development.\cite{iwata2024development} Our study points to negative viscosity as a new means of assessing cell stresses and tissue shape changes in these various applications.

\section*{IV. METHODS}

\subsection*{A. Cell culture}
Madin-Darby Canine Kidney (MDCK) type II cells were maintained in low-glucose Dulbecco's modified Eagle's medium (DMEM, 10-014, Corning) supplemented with 10\% fetal bovine serum (FBS, Corning) and 1\% Penicillin-Streptomycin (Corning). Human keratinocytes (HaCaTs) were maintained in DMEM (10-013, Corning) supplemented with 10\% FBS (Corning) and 1\% Penicillin-Streptomycin (Corning). Cells were maintained at 37$^{\circ}$C and 5\% CO$_2$. Unless otherwise stated, cells were moved to medium containing 2\% FBS 8 hr before the start of the experiment.

\subsection*{B. Chemical treatments}
Cells were treated with 0.05 {\textmu}M cytochalasin D (Sigma) and 2 {\textmu}g/mL CN03 (Cytoskeleton Inc.). Stock solutions were made by dissolving cytochalasin D in dimethylsulfoxide and CN03 in water, as in our previous work.\cite{saraswathibhatla2021}. The stock solutions were diluted in 1$\times$ phosphate buffered saline to the desired concentrations. Medium containing 10\% FBS was used for the cytochalasin D experiments (for both the control and treated groups) in order to establish a higher baseline of cellular contraction, which ensured that the decrease in contraction upon treatment would be measurable. 

\subsection*{C. Cell seeding and optical microscopy}
Polyacrylamide (PA) gels of 6 kPa Young's modulus embedded with fluorescent particles and patterned with 1 mm diameter circles were prepared using methods previously described.\cite{saraswathibhatla2020prx, notbohm2016} Briefly, 1 mm circular holes were punched into polydimethylsiloxane sheets, which were adhered to the PA gels. The holes were then coated in 0.1 mg/mL type I rat tail collagen (BD Biosciences) with the covalent crosslinker sulfo-SANPAH (50 mg/mL, Pierce Biotechnology). MDCK cells were seeded onto the masks, allowed to adhere for 1 hr, and the masks were removed. Cells were cultured to confluence overnight. For time lapse imaging, cell islands and fluorescent particles were imaged every 15 min for 15 hr. For metabolism (TMRE, 2-NBDG), and metabolic inhibition imaging, cells and fluorescent particles were imaged every 10 min for 1 or 2 hr, respectively. All imaging was performed on an Eclipse Ti-E microscope (Nikon Instruments) with a 10$\times$ numerical aperture 0.5 objective (Nikon) and an Orca Flash 4.0 digital camera (Hamamatsu) using Elements Ar software (Nikon). Imaging was performed at 37$^{\circ}$C and 5\% CO$_2$. After imaging, cells were removed by incubating in 0.05\% trypsin, and images of a traction-free reference state were collected. 

\subsection*{D. Image correlation, traction force microscopy, monolayer stress microscopy, and cell aspect ratio}

To compute cell-induced substrate displacements, Fast Iterative Digital Image Correlation (FIDIC)\cite{barkochba2015} was performed using $64 \times 64$ pixel subsets, with a spacing of 16 pixels (10.4 {\textmu}m). Cell-substrate tractions were computed using Fourier Transform Traction Cytometry\cite{butler2002} with corrections for the finite substrate thickness.\cite{delalamo2007,trepat2009} Stresses within the monolayer were computed using Monolayer Stress Microscopy \cite{tambe2011, tambe2013, saraswathibhatla2020sciData} which employs the principle of force equilibrium to the cell-substrate tractions, yielding the stress tensor in the plane of the cell layer. riefly, the method solves the equilibrium equation $\nabla \cdot \bm{\sigma} + \vec{t}/h = 0$, where $\vec{t}$ is the vector of tractions applied by the substrate to the cells, $h$ is the thickness of the cell layer (assumed to be 5 {\textmu}m), and $\bm{\sigma}$ is the stress tensor. The equilibrium equation is combined with one more equation that enforces smoothness over space to solve for the components of the stress tensor. Full details of our implementation are presented in our prior paper,\cite{saraswathibhatla2020sciData} and software used for the computation is available as described in the Code Availability section. The maximal and minimal eigenvalues of the stress tensor are the first and second principal stresses, $\sigma_1$ and $\sigma_2$, respectively, in which tensile (pulling) stress is positive, and compressive (pushing) stress is negative.

To compute cell velocities, $\Vec{v}$, we first used FIDIC to compute displacements between consecutive phase contrast images of the cells. The velocities were calculated by dividing the displacement by the time between images, which was 15 min unless otherwise stated in the Results section. The strain rate tensor $\dot{\bm{\varepsilon}}$ was calculated according to $\dot{\bm{\varepsilon}} = [\nabla \Vec{v} + (\nabla \Vec{v})^T]/2$ with $\nabla \Vec{v}$ being the velocity gradient tensor and $(\cdot)^T$ indicating the transpose. The maximal and minimal eigenvalues of the strain rate tensor are the first and second principal strain rates, $\dot{\varepsilon}_1$ and $\dot{\varepsilon}_2$, respectively. 

The shear stress was computed by $(\sigma_1-\sigma_2)/2$, and the shear strain rate was computed by $(\dot{\varepsilon}_1 - \dot{\varepsilon}_2)/2$. To compute the viscosity within local regions, a moving window was utilized . Within each window, the ratio of shear stress to the shear strain rate was taken, and the slope, computed by linear mean square regression, was used as the effective viscosity. In considering the window size to use, we aimed to choose a sufficiently large window to satisfy treating the monolayer as a continuum while maintaining a small enough window to discern spatial patterns. To guide this decision, we computed the standard deviation of the effective viscosity across space for window sizes ranging from $31\times31$ {\textmu}m$^2$ to $83\times83$ {\textmu}m$^2$ for multiple different cell islands (Appendix D, Fig. 6m). The results showed a high standard deviation for the smallest window sizes, likely due to noise. Increasing the window size from $31\times31$ {\textmu}m$^2$ to $62\times62$ {\textmu}m$^2$ substantially reduced the standard deviation; further increasing the window size led to more modest reductions in standard deviation. We chose to use a window size of $62\times62$ {\textmu}m$^2$, which is large enough to have low noise and small enough to achieve good spatial resolution.

Cell orientation was computed using the ImageJ plugin OrientationJ \cite{vromans_2016} on phase contrast images of the monolayer. A window size of 16 pixels (10.4 {\textmu}m) was used, and the ``Cubic Spline'' option for computing the gradient was used. Elongated cells were identified by taking regions in the top 10\% of the coherency.

The average viscosity lifetime was defined as the time over which the sign of effective viscosity of a given grid point remained unchanged. This was computed over all space and time.

\subsection*{E. Spatial correlations}

Spatial correlations of scalar quantities $u$ were computed according to 
\begin{equation}
    C(r\,') = \frac{\sum \Bar{u}(\Vec{r})\Bar{u}(\Vec{r}+\Vec{r}\,')}{\sum \Bar{u}^2(\Vec{r})}
\end{equation}
where $C$ is the spatial correlation, $\Bar{u} = u - u_m$ where $u_m$ is the mean of $u$, $\Vec{r}$ and $\Vec{r}\,'$ are position vectors, and $r\,' = |\Vec{r}\,'|$. The sums were calculated over all positions, $\Vec{r}$. Correlation lengths, $R$, were extracted from fitting autocorrelation curves to the exponential function $C = Ae^{-r\,'/R}$, where $C$ is the spatial autocorrelation and $A$ is a fitting parameter.

\subsection*{F. Metabolic imaging and inhibition}
Mitochondrial membrane potential was measured by incubating the cell islands for 30 min in medium (DMEM supplemented with 2\% FBS, as stated above) containing 25 {\textmu}M TMRE (Thermo). After incubation, the medium was replaced with TMRE-free medium, and the islands were imaged. 
Glucose uptake was measured by incubating cell islands for 1 hr in 400 {\textmu}M 2-NBDG (Thermo) suspended in medium that had no FBS or glucose (DMEM, no glucose, Thermo). Following the incubation, the medium was replaced with medium having no 2-NBDG, FBS, or glucose, and the islands were imaged. 
Due to the tendency for these dyes to photobleach, the ImageJ ``Bleach Correction'' function was used to correct images, using the ``histogram matching'' method.  

Metabolic inhibition experiments were performed on cell islands in a solution containing 135 mM NaCl, 5 mM KCl, 1.5 mM KCl$_2$, 1 mM MgSO$_4$, 10 mM HEPES, and 5.5 mM glucose. Metabolic inhibition was initiated with a spike of 10 mM 2-Deoxy-D-glucose (2-DG), and 2.5 mM NaCN to the solution.\cite{baron2005,smets2003}

\subsection*{G. Statistical testing}
Statistical analysis between groups was done using a Student's t-test and statistical analysis within the same group was done using a one-sample t-test. For all such tests, a sample size of at least $n = 5$ was used. Comparisons between more than two groups used one-way ANOVA followed by a Tukey Honest Significant Difference test for pair-wise comparisons. The two-sample Kolmogorov-Smirnov test was used to compare distributions. A value of $p < 0.05$ was considered statistically significant, with *, **, and *** indicating $p < 0.05$, $p < 0.01$, and $p < 0.001$, respectively. For the linear regression analyses in Fig. 1e,f, a least-squares fit was applied to the data using the function ``fitlm'' in MATLAB. The slope was reported along with the standard error, which was calculated by taking the variance of the residuals.

\section*{Acknowledgments}
We thank Christian Franck, Melissa Skala, Saverio Spagnolie, and Thomas Chandler for valuable suggestions and feedback. This work was supported by NSF grant no. CMMI-2205141 and NIH grant no. R35GM151171.

\section*{Code Availability}
Code used to analyze experimental data is available at https://github.com/jknotbohm/FIDIC and\newline 
https://github.com/jknotbohm/Cell-Traction-Stress.

\subsection*{Ethics declarations}
The authors declare no competing interests.

\newpage
\section*{Appendices}

\subsection*{Appendix A: Ruling out quantifying effective viscosity based on direct comparisons between shearing stresses and strain rates.}

Following the observation in Fig. 2e, that alignment between the first principal strain rate and second principal stress was stronger in regions of positive viscosity compared to negative, we considered whether we should have used a different method to compute the effective viscosity. Considering the prior studies that showed a reduction in effective viscosity can occur when the active stresses align with the flow,\cite{hatwalne2004, cates2008, saintillan2018} we used the stress transformation equations of continuum mechanics to project the components of the stress tensor into the orientation of the principal strain rates (Appendix D, Fig. 10a). In this orientation, there are two normal stresses, one aligned with $\dot{\varepsilon}_1$ and the second aligned with $\dot{\varepsilon}_2$, which we refer to as $\sigma_{\dot{\varepsilon}_1}$ and $\sigma_{\dot{\varepsilon}_2}$, respectively. In this frame of reference, there is also a nonzero shearing component, which, when the stress tensor is written as a symmetric matrix, is the term written off the diagonal of the matrix. Given that the definition of $\sigma_{\dot{\varepsilon}_1}$ and $\sigma_{\dot{\varepsilon}_2}$ means that they are exactly aligned with $\dot{\varepsilon}_1$ and $\dot{\varepsilon}_2$, we considered defining the effective viscosity as the ratio of $(\sigma_{\dot{\varepsilon}_1} -\sigma_{\dot{\varepsilon}_2})/2$ and $(\dot{\varepsilon}_1 - \dot{\varepsilon}_2)/2$. A representative map of this quantity showed both positive and negative regions (Appendix D, Fig. 10b). However, when the fluorescent images of cell metabolism were compared in regions where this quantity was negative to regions where this quantity was positive, the results were not statistically different (Appendix D, Fig. 10c). Hence, the ratio of $(\sigma_{\dot{\varepsilon}_1} -\sigma_{\dot{\varepsilon}_2})/2$ and $(\dot{\varepsilon}_1 - \dot{\varepsilon}_2)/2$ is not a meaningful indicator of energy injection. By contrast, using the method described in the main text, negative values of effective viscosity indicate energy injection (Fig. 4) and positive values of effective viscosity indicate viscous dissipation.\cite{mccord_positiveviscosity} In summary, the method described in the main text is a physically meaningful indicator of effective viscosity and the alternative method based on the ratio of $(\sigma_{\dot{\varepsilon}_1} -\sigma_{\dot{\varepsilon}_2})/2$ and $(\dot{\varepsilon}_1 - \dot{\varepsilon}_2)/2$ is not.

\subsection*{Appendix B: Cell-substrate tractions in regions of positive and negative viscosity.}

In considering that negative viscosity indicates energy injection into the flow via stresses within the monolayer, we considered a second potential source for energy injection, namely the tractions at the cell-substrate interface, which can have strong effects on the collective motion, such as by tuning the overall fluidity of the cell layer.\cite{bi2016, saraswathibhatla2020prx} One possibility is that, because negative viscosity indicates energy injection by active shearing stresses, less energy injection by traction is required, in which case tractions would be smaller in regions of negative viscosity. To test this possibility, we quantified the root-mean-square of tractions in regions of negative and positive viscosity and computed the ratio for multiple different cell islands. In control conditions and islands treated with cytochalasin D, the ratio was not statistically different from unity. In cell islands treated with CN03, the ratio was slightly above unity, indicating that in some cases, tractions were elevated, rather than reduced, in regions of negative viscosity (Appendix D, Fig. \ref{traction_velocity}a). Next, we considered that for the cell-substrate traction to inject energy, the traction (applied by the substrate to the cells) would have to point in the same direction as the cell velocity. We, therefore, measured the angle $\phi$ between substrate-to-cell traction and velocity in multiple cell islands. Distributions were broad with peaks at both 0$^\circ$ and 180$^\circ$ (Appendix D, Fig. \ref{traction_velocity}). The peaks at 0$^\circ$ indicate alignment between traction and velocity, meaning that, for those cells, traction injected energy into the system. Interestingly, the peak at 0$^\circ$ was more pronounced for cells exhibiting negative viscosity compared to positive (Appendix D, Fig. \ref{traction_velocity}), indicating that cells with negative viscosity inject greater energy through tractions than those with positive viscosity.  These data rule out the idea that tractions inject less energy in regions of negative viscosity.

\subsection*{Appendix C: Cell-substrate tractions in regions of high and low metabolism.}

The increased metabolic activity in regions of negative viscosity suggests that cells having surplus energy available inject it into the flow via active shearing stresses. We also considered that cells with excess energy could inject it into the flow through the cell-substrate tractions. For the cell tractions to propel the flow, the tractions applied by the substrate to the cell would have to point in the same direction as the cell's velocity. With this in mind, we calculated angle $\phi$ between traction and velocity. The metabolic activity was no different in regions of propulsive traction ($\phi < 90^{\circ}$) or resisting traction ($\phi > 90^{\circ}$) (Appendix D, Fig. \ref{metabolism_phi}), indicating that the cell-substrate tractions are not a means for injection of surplus energy into the flow.

\subsection*{Appendix D: Supplemental Figures}

This appendix contains additional Figs. 5-14.

\begin{figure}[H]
\centering
\includegraphics[keepaspectratio=true, width=3.2in]{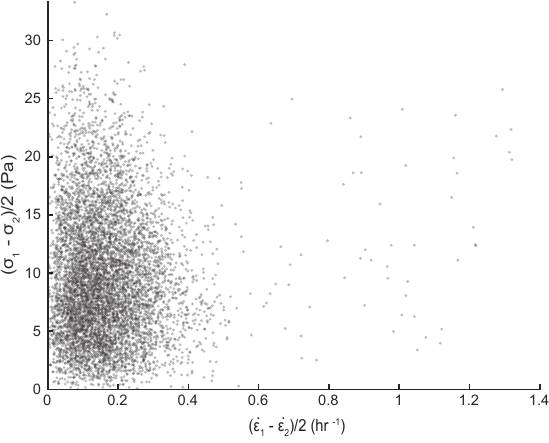}
\caption{Shear stress and shear strain rate across the entire island is uncorrelated. The representative scatter plot shows $(\sigma_{1}-\sigma_{2})/2$ against $(\dot{\varepsilon}_1-\dot{\varepsilon}_2)/2$ across an entire cell island at one point in time.}
\label{shear_island}
\end{figure}

\begin{figure}[H]
\centering
\includegraphics[keepaspectratio=true, width=6.0in]{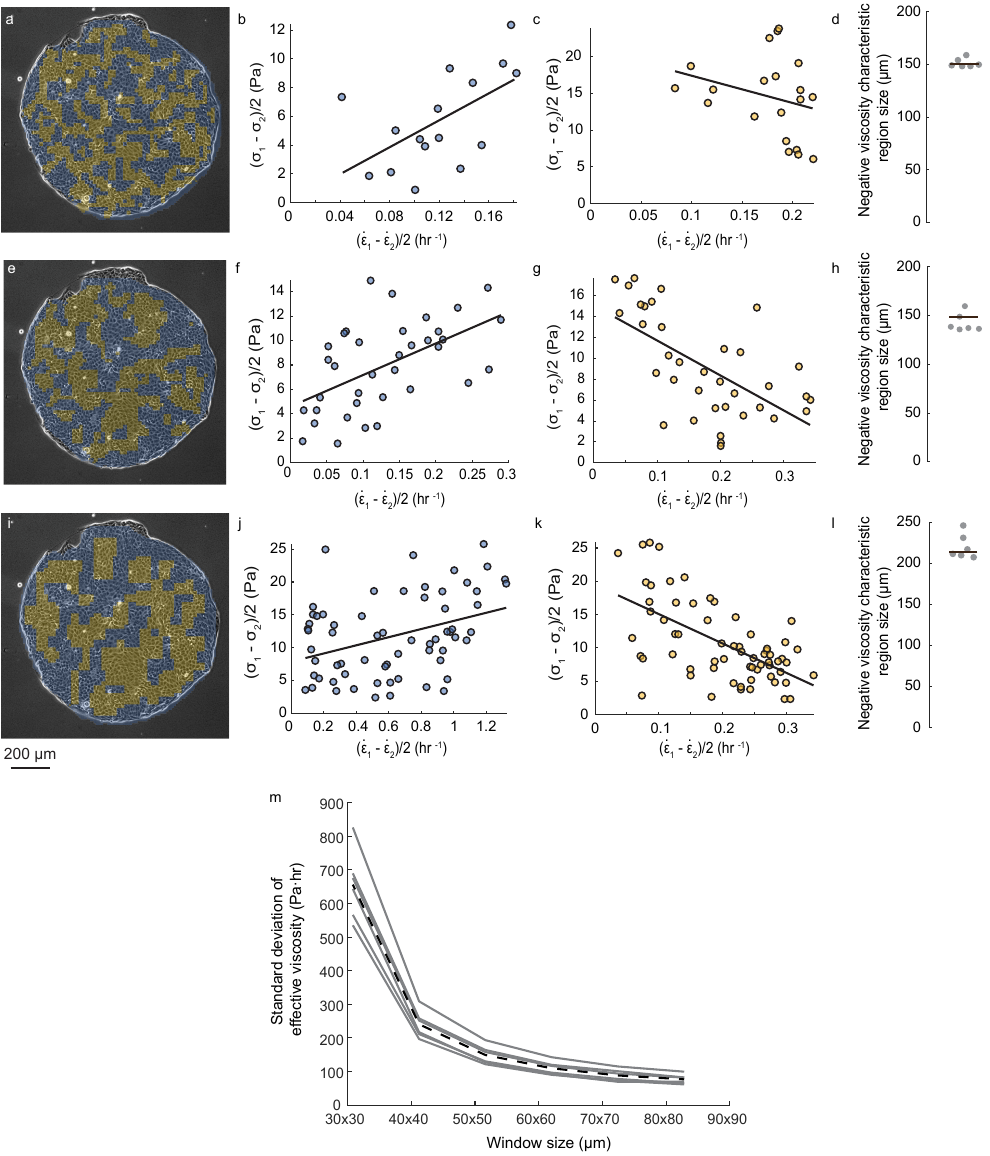}
\caption{Observation of negative viscosity is insensitive to window size.
(a) Phase contrast image of a cell monolayer overlaid with colors indicating regions of positive (blue) and negative (yellow) viscosity for a $42 \times 42$ {\textmu}m$^2$ window. 
(b) Scatter plot of shear stress and shear strain rate within a representative $42 \times 42$ {\textmu}m$^2$ window showing positive viscosity. 
(c) Scatter plot of shear stress and shear strain rate within a representative $42 \times 42$ {\textmu}m$^2$ window showing negative viscosity.
(d) Characteristic size of regions of negative viscosity using $42 \times 42$ {\textmu}m$^2$ windows. The dots represent different cell islands and the black bar indicates the mean.
Panels (e-h) and (i-l) repeat the analysis for representative $62 \times 62$ {\textmu}m$^2$ and representative $84 \times 84$ {\textmu}m$^2$ windows, respectively. 
The fact that the average size of a region of negative viscosity is similar for $42 \times 42$ {\textmu}m$^2$ and $62 \times 62$ {\textmu}m$^2$ windows indicates that the use of the $62 \times 62$ {\textmu}m$^2$ window, as in the main text, does not substantially smooth the data.
(m) The standard deviation of effective viscosity over space as a function of window size. Solid lines represent different cell islands and the dashed line indicates the mean of all solid lines. For the smallest window sizes, the standard deviation is large, likely due to large noise. The apparent noise decreased nonlinearly with increasing window size and began to plateau for window sizes greater than $52 \times 52$ {\textmu}m$^2$, suggesting window sizes slightly greater than $52 \times 52$ {\textmu}m$^2$ are a good balance between noise and spatial resolution.}
\label{window}
\end{figure}

\begin{figure}[H]
\centering
\includegraphics[keepaspectratio=true, width=5.5in]{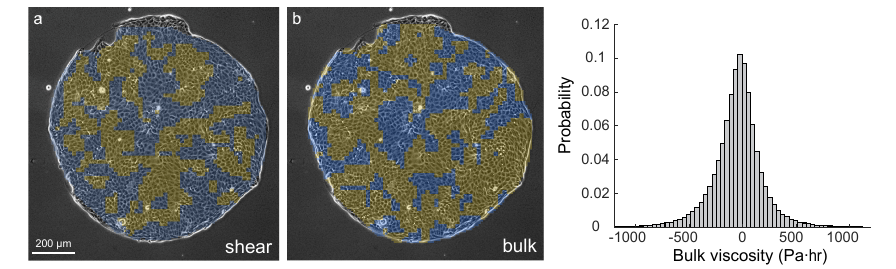}
\caption{Bulk viscosity in the cell monolayer. 
(a) Image of a cell island overlaid with regions of positive (blue) and negative (yellow) shear viscosity. 
(b) Image of a cell island overlaid with regions of positive (blue) and negative (yellow) bulk viscosity. 
(c) Histogram showing distribution of values of bulk viscosity from 6 cell islands over 15 hr of imaging.}
\label{bulk_viscosity}
\end{figure}

\begin{figure}[H]
\centering
\includegraphics[keepaspectratio=true, width=6in]{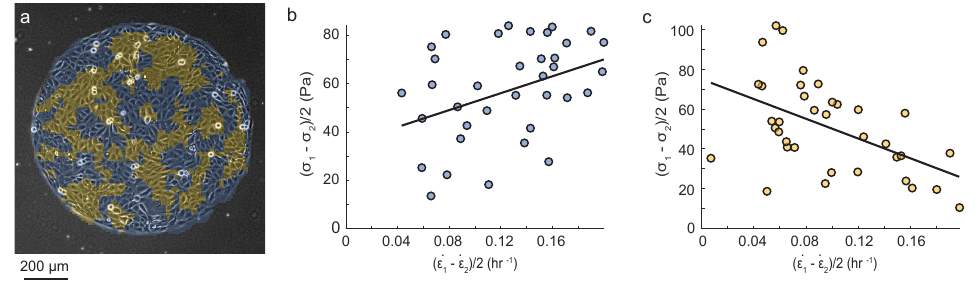}
\caption{Negative viscosity is present in multiple cell types. (a) Phase contrast image of a HaCaT cell island overlaid with regions of positive (blue) and negative (yellow) viscosity. 
(b) Scatter plot of shear stress and shear strain rate within a representative window showing positive viscosity. 
(c) Scatter plot of shear stress and shear strain rate within a representative window showing negative viscosity.}
\label{hacat}
\end{figure}

\begin{figure}[H]
\centering
\includegraphics[keepaspectratio=true, width=6.5in]{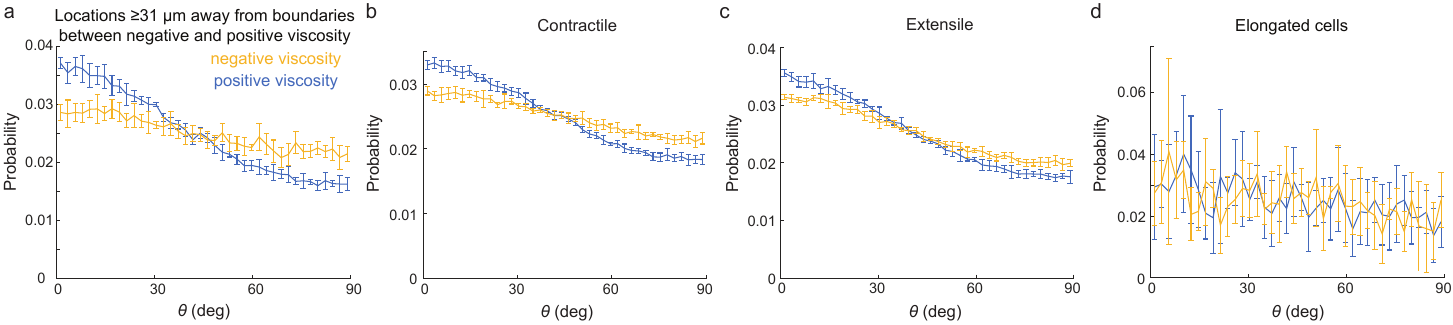}
\caption{(a) Distribution of angle $\theta$ for locations $\ge 31$ {\textmu}m away from boundaries between regions of positive and negative viscosity.
(b, c) Distributions of angle $\theta$ for contractile and extensile cells. Contractile and extensile regions in the cell layer were determined as described previously.\cite{nejad2024} The histograms show the distribution of $\theta$, the angle between orientations of $\sigma_2$ and $\dot{\varepsilon}_1$, in contractile (b) and extensile (c) regions for regions of positive and negative viscosity. 
(d) Distribution of angle $\theta$ for elongated cells.
All histograms show the distributions of $\theta$ from 6 cell islands over 15 hr of imaging.  Error bars represent the standard deviation of the probability values for each bin, calculated across all cell islands.}
\label{extensile_contractile}
\end{figure}

\begin{figure}[H]
\centering
\includegraphics[keepaspectratio=true, width=6.5in]{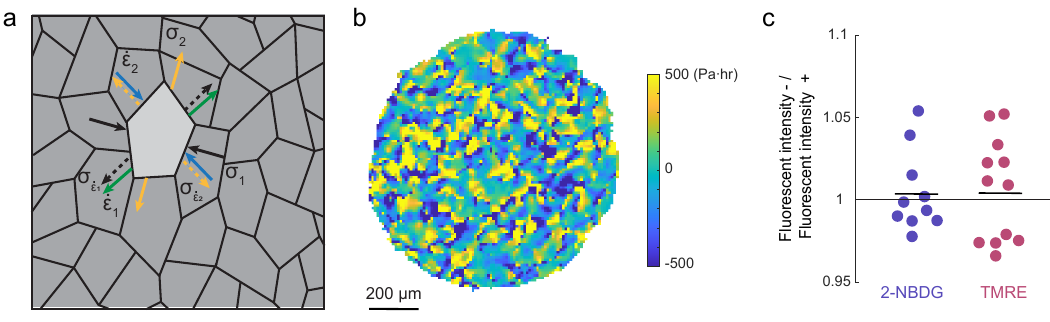}
\caption{Direct comparisons between shearing stresses and strain rates. (a) Schematic of projecting the components of the stress tensor into the orientation of the principal strain rates. The orientation of first and second principal stresses are given by $\sigma_1$ and $\sigma_2$. The orientation of first and second principal strain rates are given by $\dot{\varepsilon}_1$ and $\dot{\varepsilon}_2$. The components of the stress tensor were projected into the orientation of the principal strain rates, yielding two normal stresses, one aligned with $\dot{\varepsilon}_1$ and one aligned with $\dot{\varepsilon}_2$. We call these $\sigma_{\dot{\varepsilon}_1}$ and $\sigma_{\dot{\varepsilon}_2}$, respectively. (b) Ratio $(\sigma_{\dot{\varepsilon}_1} -\sigma_{\dot{\varepsilon}_2}) /
(\dot{\varepsilon}_1 - \dot{\varepsilon}_2)$ over space. (c) Ratio of 2-NBDG and TMRE fluorescent intensity in regions of negative $(\sigma_{\dot{\varepsilon}_1} -\sigma_{\dot{\varepsilon}_2}) /
(\dot{\varepsilon}_1 - \dot{\varepsilon}_2)$ to regions of positive $(\sigma_{\dot{\varepsilon}_1} -\sigma_{\dot{\varepsilon}_2}) /
(\dot{\varepsilon}_1 - \dot{\varepsilon}_2)$. In both cases, the data are not statistically different from unity.}
\label{aligned_visc_calc}
\end{figure}

\begin{figure}[H]
\centering
\includegraphics[keepaspectratio=true, width=3.8in]{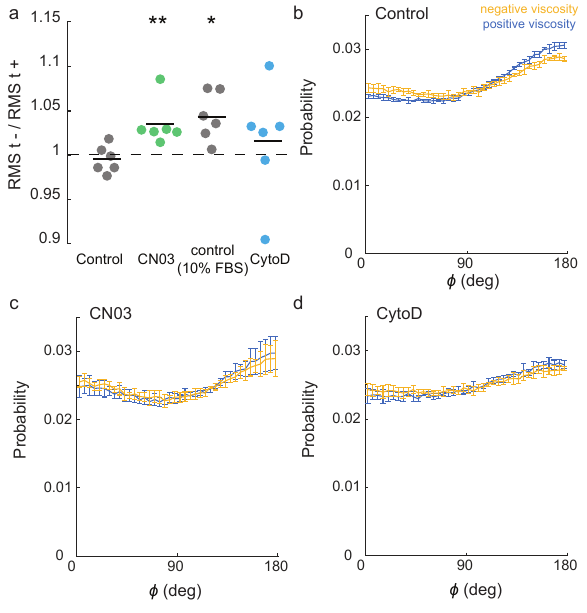}
\caption{Tractions in regions of positive and negative viscosity.
(a) Ratio of RMS traction in regions of negative viscosity to regions of positive viscosity for the control, CN03, 10$\%$ FBS control, and CytoD treated groups (Control $p = 0.058$, CN03 $p = 0.004$, Control 10$\%$ FBS $p = 0.012$, CytoD $p = 0.234$, one-sample t-test in comparison to 1). 
A dot indicates the mean of a cell island over time and black bars indicate means. 
(b--d) The angle between substrate-to-cell traction and velocity, $\phi$, in regions of positive and negative viscosity for the control case and treatments with CN03 and CytoD. Histograms show the distributions of $\phi$ from 6 cell islands over 15 hr of imaging.  Error bars represent the standard deviation of the probability values for each bin, calculated across all cell islands. The slight peak at $\phi=180^\circ$ indicates a slight tendency for traction (applied by the substrate to the cells) to be oriented opposite to the direction of velocity, thereby resisting motion within the monolayer, which is consistent with prior observations.\cite{notbohm2016}}
\label{traction_velocity}
\end{figure}

\begin{figure}[H]
\centering
\includegraphics[keepaspectratio=true, width=3.8in]{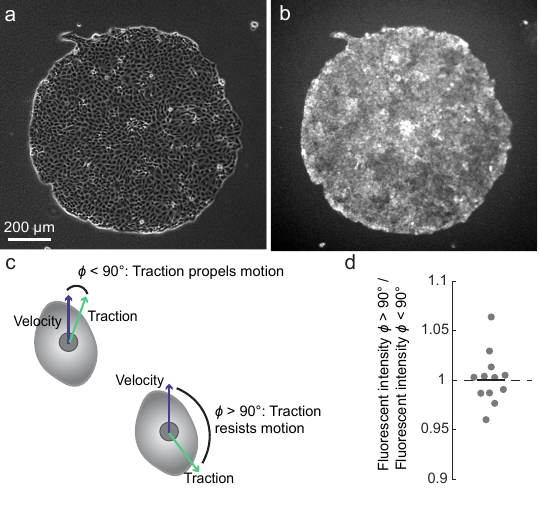}
\caption{Mitochondrial membrane potential in locations where substrate-to-cell tractions propel and resist the flow.
(a) Phase contrast image of a cell island. 
(b) Image of TMRE intensity, showing mitochondrial membrane potential. 
(c) Cartoon depicting cells with traction (applied by the substrate to the cells) and velocity aligned (left, $\phi<90^{\circ}$), and traction and velocity are misaligned (right, $\phi>90^{\circ}$). As shown in the figure, $\phi<90^{\circ}$ indicates that the tractions propel the motion, and $\phi>90^{\circ}$ indicates that the tractions resist the motion. 
(d) Ratio of fluorescent intensity in regions where velocity and traction are misaligned ($\phi > 90^\circ$) to regions where velocity and traction are aligned ($\phi < 90^\circ$). The ratio not statistically different from 1 ($p = 0.787$, one-sample t-test in comparison to 1; each dot represents an independent cell island), indicating that cellular metabolism does not correlate with whether the tractions are propulsive or resistive.}
\label{metabolism_phi}
\end{figure}

\begin{figure}[H]
\centering
\includegraphics[keepaspectratio=true, width=4.7in]{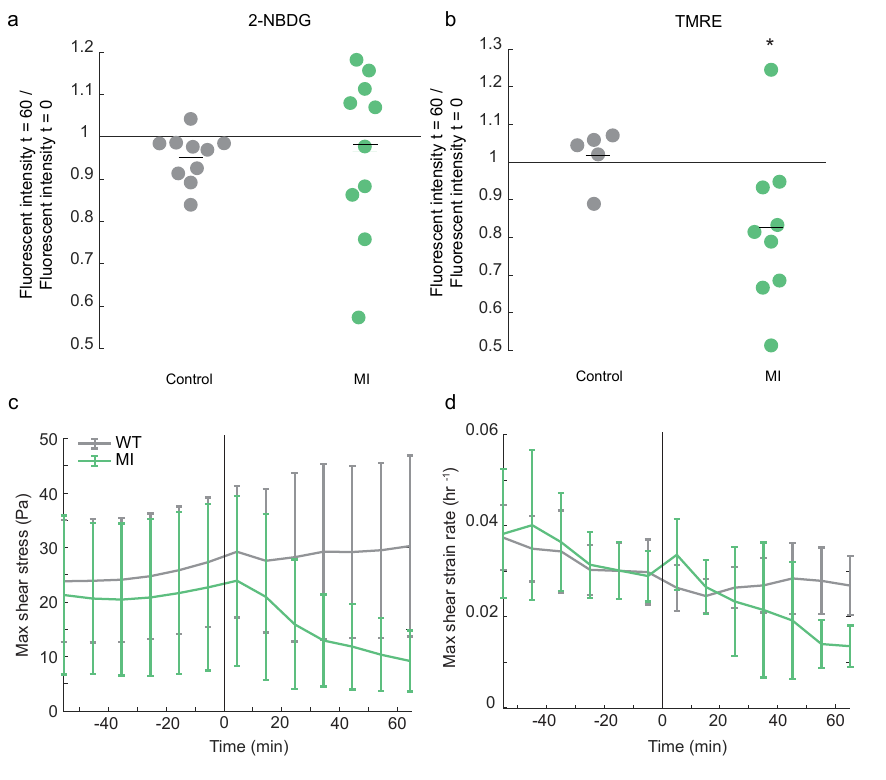}
\caption{Effects of metabolic inhibition.
(a, b) Ratio of fluorescent intensity 60 min after ($t=60$ min) to immediately before ($t=0$) spike with vehicle control and metabolic inhibitors, for 2-NBDG (glucose uptake) and TMRE (mitochondrial membrane potential). 
Metabolic inhibition caused a decrease in TMRE fluorescent intensity compared to control, indicating a decrease in mitochondrial function (TMRE, Control $p = 0.617$, MI $p = 0.020$ compared to 1, one-sample t-test). 
(c, d) Shear stress and shear strain rate over time for control (gray) and metabolic inhibition (green) cases. The metabolic inhibitors (or vehicle control) were added at time 0. Lines represent means and error bars represent standard deviations over different cell islands.}
\label{force_motion_metabolism_inhibition}
\end{figure}

\begin{figure}[H]
\centering
\includegraphics[keepaspectratio=true, width=4.7in]{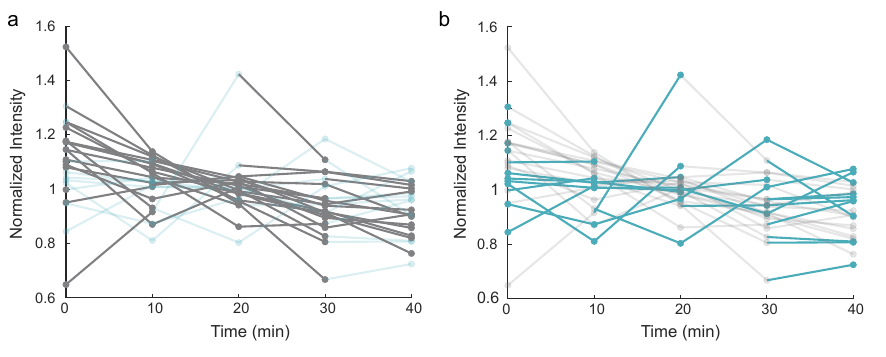}
\caption{Fluorescent intensity of TMRE over time. Each line indicates
a randomly chosen $62 \times 62$ {\textmu}m$^2$ ROI. Gray and blue line segments indicate times over which the ROI has positive and negative viscosity, respectively. Fluorescent intensity in each ROI was normalized by the mean over all time points in that ROI. These data are the same as shown as in Fig. 4i, and, for visual clarity, line transparency has been adjusted: (a) negative viscosity lines are more transparent, and (b) positive viscosity lines are more transparent. In panel a, gray lines (corresponding to positive viscosity) typically have negative slope, whereas in panel b, blue lines (corresponding to negative viscosity) typically have a slope of approximately 0. }
\label{4i_split}
\end{figure}

\newpage
\bibliographystyle{unsrt-modified}
\bibliography{refs_viscosity}

\end{document}